\documentclass{article}
\PassOptionsToPackage{table,xcdraw}{xcolor}
\usepackage{xcolor}
\usepackage{bbm}
\usepackage{multirow}
\usepackage[english]{babel}

\usepackage[letterpaper,top=2cm,bottom=2cm,left=3cm,right=3cm,marginparwidth=1.75cm]{geometry}

\usepackage{amsmath}
\usepackage{graphicx}
\usepackage[colorlinks=true, allcolors=blue]{hyperref}
\usepackage{booktabs}
\usepackage{float}
\usepackage{caption}
\newcommand{\eat}[1]{}

\usepackage{graphicx}
\usepackage{amssymb}
\usepackage{multirow}

\usepackage{caption}
\usepackage{subcaption}

\usepackage{amsmath}
\usepackage{amsthm}
\usepackage{thmtools}

\theoremstyle{definition}

\usepackage{algorithm}
\usepackage{algpseudocode}

\usepackage{array}

\usepackage[authoryear]{natbib}

\usepackage{xspace}
\usepackage[inline]{enumitem}

\usepackage{listings}
\lstdefinestyle{sqlstyle}{
  language=SQL,
  basicstyle=\footnotesize\ttfamily,
  keywordstyle=\color{blue!70!black}\bfseries,
  commentstyle=\color{gray},
  stringstyle=\color{red!60!black},
  breaklines=true,
  frame=single,
  numbers=none,
  xleftmargin=2pt,
  xrightmargin=2pt,
}
\lstdefinestyle{pystyle}{
  language=Python,
  basicstyle=\footnotesize\ttfamily,
  keywordstyle=\color{blue!70!black}\bfseries,
  commentstyle=\color{gray!70},
  stringstyle=\color{red!60!black},
  breaklines=true,
  frame=single,
  numbers=none,
  xleftmargin=2pt,
  xrightmargin=2pt,
}

\definecolor{highcol}{HTML}{1f4e79}
\definecolor{medcol}{HTML}{2e75b6}
\definecolor{ucol}{HTML}{bfbfbf}

\newcommand{\sys}{Operator Profiler\xspace}
\newcommand{\nsys}{\texttt{nsys}\xspace}
\newcommand{\ncu}{\texttt{ncu}\xspace}

\newcommand{\figref}[1]{Figure~\ref{#1}}
\newcommand{\tabref}[1]{Table~\ref{#1}}
\newcommand{\secref}[1]{Section~\ref{#1}}

\newcommand{\appref}[1]{Appendix~\ref{#1}}
\newcommand{\equref}[1]{Eq.~(\ref{#1})}
\newcommand{\etal}{\textit{et~al.}\xspace}
\newcommand{\eg}{\textit{e.g.,}\xspace}

\usepackage{pifont}
\usepackage{microtype}
\usepackage{tikz}
\usepackage{pgfplots}
\pgfplotsset{compat=1.18}
\usetikzlibrary{arrows.meta,shapes,positioning,fit,backgrounds,calc,decorations.pathreplacing}

\title{Hardware-Attributed Operator Profiling for PyTorch}

\author{
Logan Chu$^{1}$ \quad
Dong Li$^{2}$ \\
\\
$^{1}$Yotta Labs \quad
$^{2}$Yotta Labs
}
\date{}

\renewcommand{\cite}[1]{\citep{#1}}

\begin{document}
\maketitle

\begin{abstract}
Framework profilers expose operator timing without hardware counters; GPU profilers expose
hardware counters without operator attribution. Bridging this gap manually is error-prone and
does not scale. We present \textbf{\sys}, a hardware attribution pipeline that automatically
links hardware metrics to PyTorch operators via three
complementary attribution paths: torch.profiler CUPTI correlation,
NVTX temporal enclosure with per-stream interval trees, and Inductor fusion-map
enrichment from debug artifacts. NVIDIA Nsight Compute (ncu) hardware counters are matched to NVIDIA Nsight Systems (nsys) kernel records
via invocation-order matching, avoiding timestamp joins across incompatible clock domains. A curated 20-counter metric set with
duration-weighted aggregation covers all hardware bottleneck axes, layer deduplication reduces \ncu replay time by a factor of $N/K$ for models with $N$ layers across $K$ unique structural classes, and
GPU clock locking controls the kernel-duration aggregates used for operator-level comparison.
On an NVIDIA RTX PRO 6000 Blackwell, \sys attributes 95--100\% of kernel
runtime for compiled workloads (GPT-2, SDPA Attention); black-box library backends
such as cuDNN RNN are correctly surfaced as ${>}$85\% unattributed rather than silently dropped.
Applied to profile-guided FX graph optimization, attributed profiles yield
\textbf{1.76$\times$--2.24$\times$} profiled-kernel-time speedups on the two
compiled optimization case studies; a third LSTM diagnostic case identifies
cuDNN re-dispatch as a structural fix rather than an FX graph rewrite.
\end{abstract}

\section{Introduction}
\label{sec:intro}

The throughput of a GPU workload is typically dominated by a small number of hardware bottleneck
classes: memory bandwidth saturation, Tensor Core under-utilization, register-pressure-limited
occupancy, and wave starvation from insufficient parallelism. The compute-bound vs.\ memory-bound
taxonomy that Williams \etal~\cite{williams2009roofline} formalized for multicore architectures
remains difficult to apply at the operator level in PyTorch, not because the hardware data is
absent, but because no tooling bridges it to the framework abstraction.

\paragraph{The attribution gap.}
Framework profilers (\texttt{torch.profiler}, \nsys)
expose operator timing without hardware counters; \ncu exposes hardware
counters without operator attribution. Bridging the two requires matching records across tools
that run in separate passes, use incompatible clock domains (CUPTI GPU timestamps vs.\ \ncu
injection-mode timestamps), and produce outputs with no shared stable key. No existing tool was designed to bridge all three (\secref{sec:related}).

\paragraph{PyTorch compilation stack.}
In PyTorch's default eager execution, each \texttt{aten::} operator dispatches individually
through the operator dispatcher at runtime. Under TorchInductor~\cite{pytorch2024}, operators
for which a hand-tuned vendor library implementation exists---such as cuBLAS for matrix
multiplication or xFormers for attention---are emitted as extern calls that
preserve this per-operator dispatch. Operators without such external counterparts---elementwise
operations, reductions, normalization, and layout transforms---are fused in groups into
Triton~\cite{triton2019} kernels, JIT-compiled to CUDA, where each kernel may represent
multiple operators and individual \texttt{aten::} dispatch does not occur at runtime.

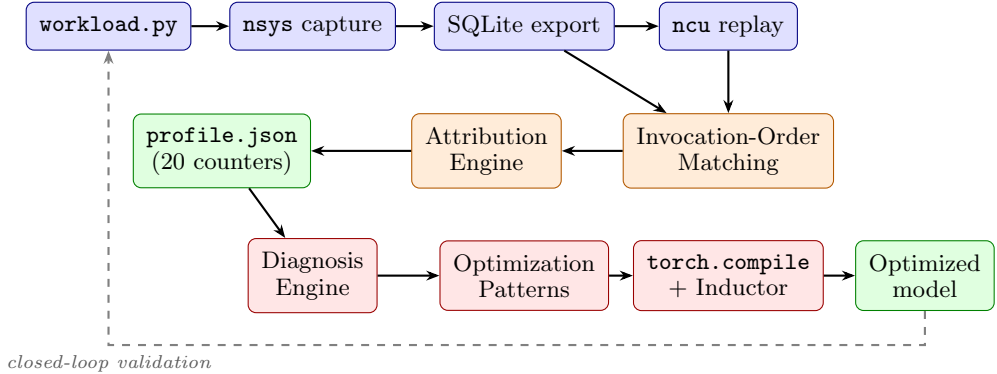
\begin{figure}[H]
\centering
\begin{tikzpicture}[
  font=\small,
  box/.style={rounded corners=3pt, draw, align=center, minimum height=1.8em, inner sep=5pt},
  toolbox/.style={box, fill=blue!12, draw=blue!50!black},
  attrbox/.style={box, fill=orange!15, draw=orange!70!black},
  outbox/.style={box, fill=green!12, draw=green!50!black},
  pgobox/.style={box, fill=red!10, draw=red!60!black},
  arr/.style={-{Stealth[length=5pt]}, thick},
  darr/.style={-{Stealth[length=5pt]}, thick, dashed, gray},
  lbl/.style={font=\scriptsize\itshape, text=gray!70!black},
]

\node[toolbox] (workload) at (0,0)   {\texttt{workload.py}};
\node[toolbox] (nsys)     at (2.7,0) {\nsys capture};
\node[toolbox] (sqlite)   at (5.5,0) {SQLite export};
\node[toolbox] (ncu)      at (8.2,0) {\ncu replay};

\draw[arr] (workload) -- (nsys);
\draw[arr] (nsys)     -- (sqlite);
\draw[arr] (sqlite)   -- (ncu);

\node[attrbox] (iom)    at (8.2,-1.65) {Invocation-Order\\[-1pt]Matching};
\node[attrbox] (sit)    at (5,-1.65) {Attribution\\[-1pt]Engine};
\node[outbox]  (prof)   at (1.5,-1.65) {\texttt{profile.json}\\[-1pt](20 counters)};

\draw[arr] (sqlite) -- (iom);
\draw[arr] (ncu)    -- (iom);
\draw[arr] (iom)    -- (sit);
\draw[arr] (sit)    -- (prof);

\node[pgobox] (diag)    at (2.7,-3.3)  {Diagnosis\\[-1pt]Engine};
\node[pgobox] (passes)  at (5.5,-3.3)  {Optimization\\[-1pt]Patterns};
\node[pgobox] (compile) at (8.2,-3.3)  {\texttt{torch.compile}\\[-1pt]+ Inductor};
\node[outbox] (opt)     at (10.8,-3.3) {Optimized\\[-1pt]model};

\draw[arr] (prof)    -- (diag);
\draw[arr] (diag)    -- (passes);
\draw[arr] (passes)  -- (compile);
\draw[arr] (compile) -- (opt);

\draw[darr] (opt.south) -- ++(0,-0.45) -| (workload.south)
  node[midway, below, lbl] {closed-loop validation };

\end{tikzpicture}
\caption{\textbf{\sys system overview.} The profiling stage (top) runs \nsys and \ncu to collect kernel timelines and hardware counters. The attribution engine (middle) combines these into \texttt{profile.json}. The optimization row (bottom) shows a downstream application of the attributed profile; it is not part of the attribution pipeline itself.}
\label{fig:overview}
\end{figure}

\paragraph{Contributions.}
\sys decomposes the attribution problem into two orthogonal steps.
First, hardware counters from \ncu are joined to kernel records in the \nsys trace via
invocation-order matching.
Second, operator identity is assigned to each counter-enriched kernel record through three
complementary attribution paths: CUPTI correlation via \texttt{torch.profiler}, NVTX temporal
enclosure, and Inductor fusion-map enrichment.
\figref{fig:overview} shows the complete end-to-end pipeline.
Applied to profile-guided FX graph optimization, attributed profiles yield
\textbf{1.76$\times$--2.24$\times$} profiled-kernel-time speedups on the two
compiled optimization case studies, while an LSTM diagnostic case demonstrates
that the UNATTRIBUTED tier can identify when the correct fix is structural
cuDNN re-dispatch rather than FX graph rewriting.

\section{Related Work}
\label{sec:related}

The tools closest to \sys fall short along three primary dimensions: operator-level timing
without counters, kernel-level counter collection without operator identity, and
compilation-level optimization without counter-verified feedback. Adjacent areas---fused kernel
libraries and profile-guided optimization---are surveyed below for completeness.

\paragraph{Operator-level profilers.}
\texttt{torch.profiler} and its Kineto backend \cite{kineto2021} record wall-clock durations and
Chrome traces at the \texttt{aten::} operator granularity~\cite{pytorch2019} and expose CUDA activity correlation
via CUPTI External IDs---giving kernel-level \emph{timing} attribution but no access to hardware
performance counters. Practitioners can identify \emph{which} operator is slow, but cannot
determine \emph{which hardware resource} it is contending for---the distinction that determines
what optimization is warranted. Without hardware
counter data at the operator level, optimization choices remain heuristic.
NVIDIA DLProf~\cite{dlprof} is the closest NVIDIA-specific predecessor: it uses NVTX markers
to correlate CPU and GPU time with model operations, reports at operation, layer, and kernel
granularity, and includes Tensor Core usage diagnostics. DLProf remains timing-centric:
its GPU-side data comes from \nsys, and Tensor Core utilization is inferred from kernel names.
\sys adds the missing counter-attribution layer. It joins application-mode \ncu hardware
counter rows to \nsys kernel records through invocation-order matching, then assigns those
counter-enriched kernel records to PyTorch operators, including Inductor-era fused kernels.
This produces per-operator hardware metrics such as measured Tensor Core activity,
occupancy, memory throughput, register pressure, and cache behavior rather than only
operation timing or kernel-name-derived Tensor Core eligibility.

\paragraph{Kernel-level profilers.}
NVIDIA Nsight Systems (\nsys) \cite{nsightsystems} captures the full CUDA kernel timeline with
NVTX annotations at high throughput, but produces no operator-level view: every kernel appears by
its internal CUDA or Triton name (\eg \texttt{cutlass\_80\_simt\_sgemm\_128x32\_tn}), with no
indication of which PyTorch operator dispatched it. NVIDIA Nsight Compute (\ncu)
\cite{nsightcompute} collects per-kernel hardware counters at full precision---occupancy, DRAM
bytes, Tensor Core utilization, cache hit rates---but its output is keyed on kernel names and
invocation indices in the replay timeline, again with no awareness of the operator graph above.
NVIDIA DCGM and Dynolog \cite{dynolog2023} operate at cluster monitoring granularity and do not
perform per-kernel or per-operator attribution. Taken together, these tools collectively address
hardware counter collection and kernel-timeline capture but leave kernel-to-operator attribution entirely unresolved.

\paragraph{Deep learning compilers and optimizers.}
TorchDynamo and TorchInductor \cite{pytorch2024} compile PyTorch programs via FX graph capture~\cite{torchfx2022} to Triton kernels
using heuristic fusion rules (pointwise chaining, matmul epilogue fusion), but fusion decisions
are made without hardware counter feedback: whether a given fusion improved Tensor Core
utilization is not a question the standard Inductor toolchain can answer programmatically. TVM \cite{tvm2018} and Halide \cite{halide2013} optimize tensor programs via
measurement-based auto-tuning and schedule/algorithm separation respectively; both optimize
\emph{latency} directly rather than using hardware counters to identify the bottleneck class
driving that latency. ONNX Runtime\footnote{\url{https://onnxruntime.ai}} and NVIDIA
TensorRT\footnote{\url{https://developer.nvidia.com/tensorrt}} perform layer fusion and
quantization behind black-box interfaces that cannot be extended for custom architectures or
augmented with attribution data.

\paragraph{Fused kernel libraries.}
FlashAttention \cite{flashattention2022,flashattention2024} and xFormers \cite{xformers2022}
provide hand-optimized fused attention kernels that eliminate DRAM round-trips by keeping
intermediate tensors in on-chip SRAM. These represent the class of optimization that
attributed hardware counter evidence should motivate rather than apply unconditionally; what
is absent is a profiling layer that identifies, from measured per-operator counters, which
variant is warranted for a given workload.

\paragraph{Profile-guided optimization.}
Classic PGO in C/C++ compilers (GCC, LLVM) collects branch frequency profiles to guide inlining
and layout decisions \cite{autofdo2016}. Applying the same principle at the GPU operator level requires hardware counter data attributed
to specific operators. To our knowledge, \sys is the first published system to combine
\nsys kernel traces, application-mode \ncu hardware counters, and PyTorch operator
attribution through invocation-order matching, and to use the resulting attributed profile
to drive hardware-informed FX graph optimization.

\section{Attribution Pipeline}
\label{sec:attribution}

The na\"{i}ve approach---run \nsys for the kernel timeline, run \ncu for hardware counters, and
join on kernel name---fails on three fronts. First, kernel names are not unique identifiers:
any kernel invoked more than once (\eg the same GEMM across 12 transformer layers) produces
multiple indistinguishable rows with no stable key. Timestamps are the natural tiebreaker,
but \nsys and \ncu use incompatible clock domains (\secref{sec:clockdomain}), making
timestamp joins systematically incorrect. Second, even a correctly joined kernel record
carries no operator identity: Inductor-compiled Triton kernels bear opaque hash suffixes
(\eg \texttt{triton\_poi\_fused\_relu\_addmm\_0}) that do not map to a single
\texttt{aten::} operator without additional artifacts. Third, manual bridging is
$O(\text{operators})$ effort per model variant and does not scale. \sys addresses all three
failures through the pipeline described below.

\begin{figure}[H]
\centering
\begin{tikzpicture}[
  font=\small, >=Stealth,
  decision/.style={diamond, draw, fill=yellow!15, align=center,
                   inner sep=2pt, aspect=2, font=\small},
  result/.style={rounded corners=4pt, draw, align=center,
                 minimum width=4.2cm, inner sep=5pt},
  high/.style={result, fill=blue!15, draw=blue!50!black},
  med/.style={result, fill=cyan!12, draw=cyan!50!black},
  unattr/.style={result, fill=gray!20, draw=gray!50},
  arr/.style={-Stealth, thick},
]
  \node[draw, rounded corners=3pt, fill=white, font=\small\bfseries,
        inner sep=5pt, align=center] (k) at (0,0) {kernel $k$\\(name, invocation index)};
  \node[decision] (d1) at (0,-2.2)
    {\texttt{corr.json} entry\\for \texttt{(name, idx)}?};
  \draw[arr] (k) -- (d1);
  \node[high] (high) at (7,-2.2)
    {\textbf{HIGH}\\torch.profiler CUPTI\\correlation};
  \draw[arr] (d1) -- node[above,font=\scriptsize]{YES} (high);
  \node[decision] (d2) at (0,-5.0)
    {NVTX enclosure at\\host launch time $T_k$?};
  \draw[arr] (d1) -- node[right,font=\scriptsize]{NO} (d2);
  \node[med] (mnvtx) at (7,-5.0)
    {\textbf{MEDIUM}\\NVTX enclosure\\(innermost range)};
  \draw[arr] (d2) -- node[above,font=\scriptsize]{YES} (mnvtx);
  \node[decision] (d3) at (0,-7.8)
    {Inductor fusion map\\entry for kernel name?};
  \draw[arr] (d2) -- node[right,font=\scriptsize]{NO} (d3);
  \node[med] (mind) at (7,-7.8)
    {\textbf{MEDIUM}\\Inductor fusion\\enrichment};
  \draw[arr] (d3) -- node[above,font=\scriptsize]{YES} (mind);
  \node[unattr] (unattr) at (0,-10.2)
    {\textbf{UNATTRIBUTED}\\cuDNN-internal; eager Triton\\without NVTX or debug output};
  \draw[arr] (d3) -- node[right,font=\scriptsize]{NO} (unattr);
\end{tikzpicture}
\caption{\textbf{Attribution priority decision tree.}
  Each kernel is tested against the three attribution paths in descending confidence
  order.  A HIGH-confidence entry from the \texttt{torch.profiler} correlation pass
  (\secref{sec:tprofiler}) pre-empts the NVTX and Inductor checks.  The Inductor
  fusion enrichment pass (\secref{sec:heuristic}) is applied as a \emph{post-attribution}
  augmentation at all tiers: it populates \texttt{fused\_ops} without overriding
  an existing attribution.  Only kernels that fail all three paths land in
  \texttt{unattributed\_kernels[]}.}
\label{fig:decisiontree}
\end{figure}
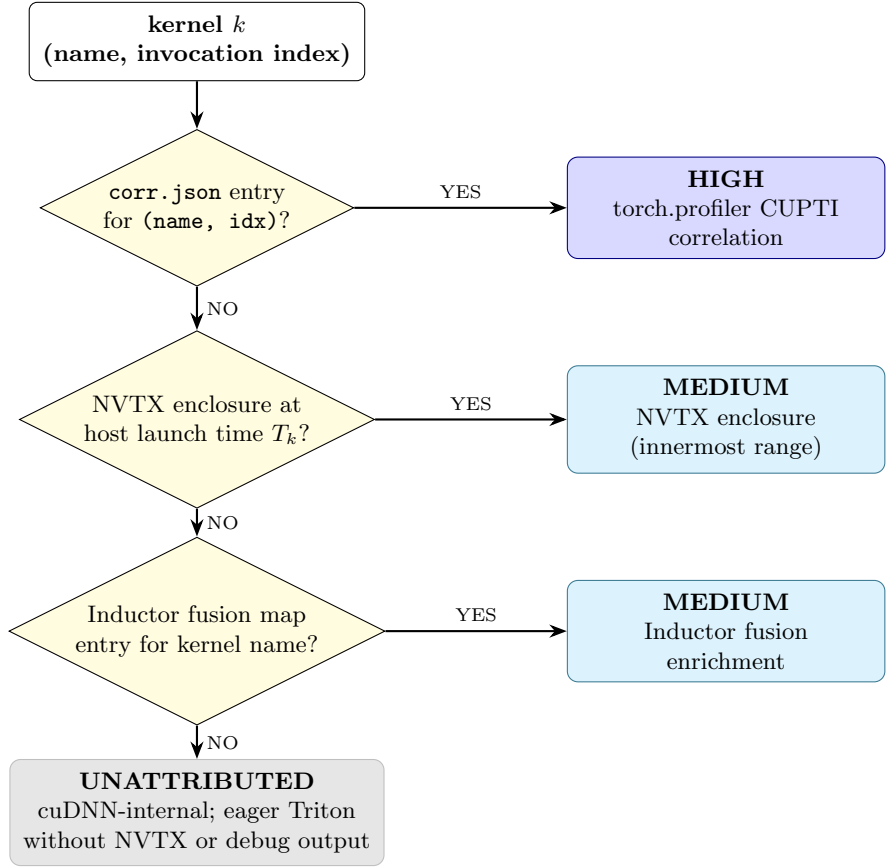

\sys ingests two data sources produced for the same workload run: (1) an \nsys SQLite export
containing CUDA kernel activity and NVTX range annotations, and (2) an \ncu CSV output containing
per-kernel hardware counter measurements. The pipeline produces a hardware-attributed
\texttt{OperatorProfile}: a JSON document mapping each dispatched PyTorch operator to its constituent CUDA kernels,
along with the hardware metrics and attribution confidence of each kernel.
For Inductor-compiled workloads, fused kernels are annotated with all constituent
\texttt{aten::} operators rather than attributed to a single dispatch call.
Attribution is assigned at one of three confidence tiers (\tabref{tab:tiers}, \figref{fig:decisiontree}); per-workload coverage fractions are reported in \secref{sec:coverage}.

\begin{table}[H]
\centering
\caption{Attribution methods in \sys.}
\label{tab:tiers}
\small
\setlength{\tabcolsep}{4pt}
\begin{tabular}{p{3.0cm}p{4.5cm}p{6.0cm}}
\toprule
\textbf{Method} & \textbf{Evidence} & \textbf{Trigger Condition} \\
\midrule
\rowcolor{highcol!15}
HIGH & torch.profiler CUPTI correlation (IOM, \secref{sec:clockdomain}) & Kineto trace available; kernel count agrees with dispatch count \\
\rowcolor{medcol!15}
MEDIUM & NVTX enclosure (interval tree) & Kernel launch falls within an \texttt{aten::} NVTX range \\
\rowcolor{medcol!10}
MEDIUM & Inductor debug artifacts & Kernel name in Inductor fusion map; NVTX enclosure absent \\
\rowcolor{ucol!15}
UNATTRIBUTED & None & cuDNN-internal kernels (LSTM, cuDNN conv.); eager Triton without NVTX \\
\bottomrule
\end{tabular}
\end{table}

\subsection{Data Ingestion}
\label{sec:ingestion}

A fundamental resource contention constrains the capture architecture: \nsys and
\texttt{torch.profiler} both register CUPTI subscribers for kernel timing and cannot coexist
within the same process. \nsys's CUPTI subscriber preempts the
\texttt{torch.profiler} callback silently; the correlation pass terminates with a
\texttt{.corr.json} containing zero entries and produces no diagnostic indication of failure.
Thus, two sequential, non-overlapping invocations are needed. The first executes as an untraced Python process to capture \texttt{torch.profiler} correlation output and the second executes under \texttt{nsys profile} and reuses the
Inductor compilation cache produced prior.

\textbf{Nsight Systems export.}
\sys extracts three classes of records from the SQLite nsys output:
GPU-domain kernel timestamps and launch geometry; the host-side API timestamp
at the moment each \texttt{cuLaunchKernel} call returned; and completed NVTX
operator range annotations.  The host and GPU records are linked by a CUPTI
correlation identifier---an integer that CUPTI assigns to both the CPU-side
launch call and the resulting GPU kernel execution, enabling a join across the
two tables in the nsys SQLite export.
The host-side launch timestamp serves as a CPU-domain anchor for NVTX bracket queries
without conflating it with the GPU-domain kernel execution timestamp (\secref{sec:clockdomain}).

\textbf{Nsight Compute export.}  Collecting hardware counters for every kernel in the \nsys
trace requires selecting an \ncu replay strategy.  Two candidates were evaluated and rejected.
The first was a per-kernel-name subprocess loop in which \ncu is invoked once per unique
kernel name with a name filter, each constituting an independent replay.  The profiling cost
is $O(\text{unique kernels})$: for a model such as GPT-2, which produces hundreds of distinct
Triton kernel variants, this approach is computationally prohibitive at scale.  The second
candidate was \ncu's \texttt{--filter-by-nvtx-range}, which confines each replay to kernels
enclosed within a specified \texttt{aten::} range.  This mode fails silently: \ncu's injection
mechanism targets the dynamically-loaded NVTX library, whereas PyTorch~2.x links NVTX
statically inside \texttt{libtorch\_cuda.so}, rendering its annotations invisible to \ncu
irrespective of \texttt{emit\_nvtx} instrumentation---and producing no diagnostic output to
indicate that the collected counters are empty.

\sys therefore uses application-mode replay, in which \ncu re-executes the entire workload
once per hardware counter group (4--8 passes total), collecting all 20 metrics simultaneously.
This reduces profiling cost to $O(\text{counter groups})$ rather than $O(\text{unique kernels})$---a
10--50$\times$ reduction---and is the only mode that reliably collects counters for
Inductor-compiled workloads.

\subsection{GPU Clock Locking for Reproducible Duration Comparisons}
\label{sec:clocklocking}

Duration fields in \texttt{profile.json} come from the \nsys capture, not from
\ncu replay.  Each kernel record stores \texttt{duration\_ns} as the CUPTI
GPU-domain interval \texttt{end\_ns - start\_ns}; each operator record stores
\texttt{aggregated.total\_duration\_ns} as the sum of those \nsys kernel durations
over kernels attributed to that operator.  The \ncu metric
\texttt{gpu\_\_time\_duration.sum} is retained only as a raw replay-phase counter
inside \texttt{metrics.raw} and is not used for speedup reporting.  These
kernel-duration aggregates are subject to a subtle confound: a GPU's boost clock
floats with die temperature and power draw, typically varying 5--15\% across
captures separated by minutes or hours.  An optimized workload dissipates less
power than the baseline (fewer FLOPs on the SIMT path), so its capture may
execute at a \emph{higher} sustained clock frequency, artificially elevating the
apparent speedup.  Conversely, thermal throttling during a prolonged baseline
capture suppresses the measured baseline duration.  Both effects bias the
$\text{baseline}/\text{optimized}$ duration ratio even when no architectural
change was made.

\textbf{Probe-and-lock.}
\sys pins the SM and memory clocks for the duration of the \nsys capture.  The default
target is \emph{probe-and-lock}: a synthetic matrix-multiply load runs for ${\sim}1.5\,\text{s}$
(with a $1\,\text{s}$ warmup to reach thermal steady state), and the median clock observed
during the probe is selected as the lock target.  Locking to the \emph{sustained} frequency
(rather than the peak boost clock, which the GPU cannot maintain under load) guarantees the
lock is genuinely held throughout the capture.

\textbf{Cache and reuse.}
The probed clock is written to a per-GPU clock cache with a 6-hour TTL.  The baseline capture probes once; the optimized capture reuses
the cached clock.  Both captures therefore execute at exactly the same frequency, making
their duration ratio clock-immune by construction---the same guarantee that invocation-order
matching provides for the hardware counter join.

\textbf{Scope.}
Clock locking applies to the \nsys capture phase only.  The \ncu replay phase is
unaffected: \ncu already self-locks to its own base clock before each replay pass, and
the hardware counter values it reports are independent of the GPU's runtime SM frequency.

\subsection{The Clock Domain Problem and Invocation-Order Matching}
\label{sec:clockdomain}

\begin{figure}[H]
\centering
\begin{tikzpicture}[font=\small, >=Stealth]
  \node[anchor=east, font=\small\bfseries] at (-0.1,3.25) {\nsys};
  \node[anchor=east, font=\scriptsize, gray] at (-0.1,2.95) {CUPTI GPU clock};
  \draw[->,thick] (0,3.1) -- (11.8,3.1);
  \draw[fill=blue!20,draw=blue!50] (0.55,2.85) rectangle (1.45,3.35);
  \node[font=\scriptsize] at (1.0,3.1) {$K_A$};
  \draw[fill=green!20,draw=green!60!black] (2.05,2.85) rectangle (2.95,3.35);
  \node[font=\scriptsize] at (2.5,3.1) {$K_B$};
  \draw[fill=cyan!20,draw=cyan!60!black] (3.55,2.85) rectangle (4.45,3.35);
  \node[font=\scriptsize] at (4.0,3.1) {$K_C$};
  \draw[fill=orange!20,draw=orange!60] (5.55,2.85) rectangle (6.45,3.35);
  \node[font=\scriptsize] at (6.0,3.1) {$K_D$};
  \draw[fill=red!15,draw=red!50] (7.55,2.85) rectangle (8.45,3.35);
  \node[font=\scriptsize] at (8.0,3.1) {$K_E$};
  \node[anchor=east, font=\small\bfseries] at (-0.1,0.9) {\ncu};
  \node[anchor=east, font=\scriptsize, gray] at (-0.1,0.6) {replay clock};
  \draw[->,thick] (0,0.75) -- (11.8,0.75);
  \draw[fill=blue!20,draw=blue!50] (0.85,0.5) rectangle (1.75,1.0);
  \node[font=\scriptsize] at (1.3,0.75) {$K_A$};
  \draw[fill=green!20,draw=green!60!black] (2.65,0.5) rectangle (3.55,1.0);
  \node[font=\scriptsize] at (3.1,0.75) {$K_B$};
  \draw[fill=cyan!20,draw=cyan!60!black] (4.45,0.5) rectangle (5.35,1.0);
  \node[font=\scriptsize] at (4.9,0.75) {$K_C$};
  \draw[fill=orange!20,draw=orange!60] (7.05,0.5) rectangle (7.95,1.0);
  \node[font=\scriptsize] at (7.5,0.75) {$K_D$};
  \draw[fill=red!15,draw=red!50] (9.55,0.5) rectangle (10.45,1.0);
  \node[font=\scriptsize] at (10.0,0.75) {$K_E$};
  \draw[green!60!black,thick] (1.0,2.85) -- (1.3,1.0);
  \draw[green!60!black,thick] (2.5,2.85) -- (3.1,1.0);
  \draw[green!60!black,thick] (4.0,2.85) -- (4.9,1.0);
  \draw[green!60!black,thick] (6.0,2.85) -- (7.5,1.0);
  \draw[green!60!black,thick] (8.0,2.85) -- (10.0,1.0);
  \draw[dashed,red!75,very thick] (6.0,2.85) -- (4.9,1.0);
  \node[font=\scriptsize,red!75,align=center] at (4.9,2.45)
    {\textit{wrong:}\\$K_D \!\leftrightarrow\! K_C$};
  \node[above,font=\scriptsize,orange!80!black] at (6.25,1.0)
    {$\Delta t$ grows};
  \draw[green!60!black,thick] (10.0,2.1) -- (10.6,2.1);
  \node[right,font=\scriptsize] at (10.6,2.1) {IOM: ordinal join};
  \draw[dashed,red!75,thick] (10.0,1.75) -- (10.6,1.75);
  \node[right,font=\scriptsize] at (10.6,1.75) {timestamp join (fails)};
\end{tikzpicture}
\caption{\textbf{Clock domain divergence and invocation-order matching.}
  The \nsys CUPTI GPU clock and the \ncu injection replay clock diverge by
  10--100\,$\mu$s per kernel due to PMU serialization overhead; the gap
  \emph{accumulates} across the trace.  A na\"{i}ve timestamp join (dashed red)
  maps the \nsys record of $K_D$ to the \ncu row for $K_C$---a wrong counter
  assignment.  Invocation-Order Matching (IOM, green) connects each kernel in the
  \nsys trace to the matching \ncu row by name and position,
  making the join clock-independent.}
\label{fig:clockdomain}
\end{figure}
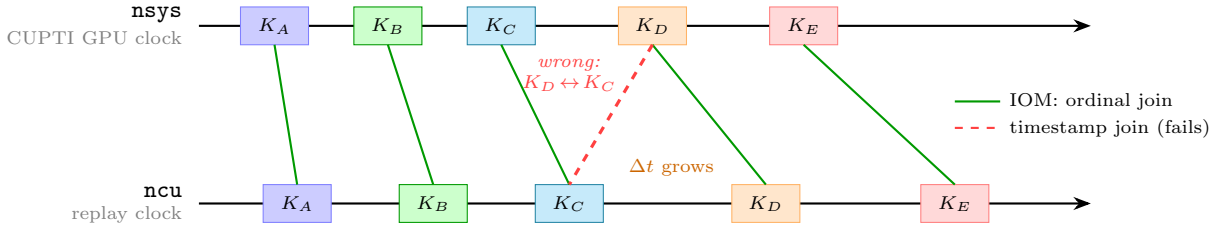

The fundamental obstacle in cross-tool profiling is that \nsys and \ncu observe the same kernel
execution through incompatible timing mechanisms. \nsys records CUPTI GPU-domain timestamps from
the hardware's free-running clock. \ncu replays kernels in an isolated profiling pass with injected
performance counter reads; its internal invocation clock reflects replay-mode execution, not the
original capture clock. On a single GPU, these two clocks routinely diverge by 10--100~$\mu$s per
kernel due to replay serialization and Performance Monitoring Unit (PMU) injection overhead (\figref{fig:clockdomain}). A na\"{i}ve timestamp join would
incorrectly associate hardware counter measurements with the wrong operator---a failure mode
particularly susceptible to misattribution for operators with similar execution times.

\textbf{Invocation-Order Matching (Algorithm~\ref{alg:iom}).}  The key insight is that \ncu
replays kernels in the same execution order they appear in the \nsys timeline: the $i$-th
invocation of kernel $K$ in the \nsys trace corresponds to the $i$-th row for kernel $K$ in the
\ncu CSV output.  We exploit this ordering to assign hardware metrics without any timestamp
comparison. The matching is correct
when the workload is \emph{deterministic}: the same kernel names must
appear in the same counts and the same execution order in both the \nsys capture and the \ncu
replay.  Concretely, models with conditional control flow, dynamic dispatch, or
allocator-emitted workspace kernels can violate this invariant without producing a shape
error.  \sys validates that per-kernel invocation counts match between the two runs before
issuing counter assignments; a count mismatch on any kernel name is logged as a
warning and metric assignment for that kernel is skipped.

\begin{algorithm}[t]
\caption{Invocation-Order Matching (\nsys $\times$ \ncu)}
\label{alg:iom}
\begin{algorithmic}[1]
\Require \texttt{nsys\_kernels}: list of \texttt{KernelRow} sorted by \texttt{start\_ns} per stream
\Require \texttt{ncu\_rows}: list of \texttt{NcuRow} sorted by invocation index per kernel name
\Ensure  Each \texttt{KernelRow} assigned a \texttt{KernelMetrics} object
\State $\textit{counter} \gets \{\,\}$
  \Comment{invocation count per kernel name, initialized to 0}
\For{$k \in \texttt{nsys\_kernels}$}
  \State $i \gets \textit{counter}[k.\textit{name}]$
  \If{$i \ge |\texttt{ncu\_rows}[k.\textit{name}]|$}
    \State \textbf{warn}($k.\textit{name}$, $i$); \textbf{continue}
    \Comment{invocation count mismatch: log warning, skip metric assignment}
  \EndIf
  \State $k.\textit{metrics} \gets \texttt{ncu\_rows}[k.\textit{name}][i]$
  \State $\textit{counter}[k.\textit{name}] \mathrel{+}= 1$
\EndFor
\end{algorithmic}
\end{algorithm}

\subsection{torch.profiler Correlation Attribution (HIGH Confidence)}
\label{sec:tprofiler}

The highest-confidence attribution path exploits CUPTI's External Activity API~\cite{cupti2024}, which records a
causal link between CPU-side operator dispatch and the GPU kernels it launches.  When
\texttt{torch.profiler} is active with the Kineto backend, each \texttt{aten::} dispatch call is
assigned a correlation ID at the moment of \texttt{cuLaunchKernel}; CUPTI propagates this ID
through the driver stack and attaches it to every kernel invocation that the dispatch triggers.
The result is a hardware-recorded mapping
$\{(\textit{kernel\_name},\ i) \to \texttt{aten::op}\}$ that requires no timestamp comparison
and no pattern matching---a causal record of dispatch rather than an inference from proximity.

\textbf{Capture and output.}  As noted in \secref{sec:ingestion}, \texttt{torch.profiler} and
\nsys cannot coexist within the same process.  The correlation pass therefore runs as a standalone
Phase~1 invocation with no \nsys, producing a correlation map from
\texttt{(kernel\_name, invocation\_index)} to \texttt{aten::op\_name}.
The attribution engine loads this map and applies it as the first-priority tier;
any kernel with a matching entry receives HIGH-confidence attribution before the NVTX and
Inductor passes are consulted.

\textbf{Validity condition.}  The same determinism invariant from \secref{sec:clockdomain}
applies across phases: \texttt{--warmup-iters} and \texttt{--measure-iters} must be identical
across Phase~1, Phase~2, and the \ncu replay, as any iteration-count difference shifts
invocation indices and corrupts the mapping.  In practice, the condition holds for all
evaluated workloads once the Inductor compilation cache is warm.

\subsection{StreamIntervalTree Attribution}
\label{sec:intervaltree}

NVTX (NVIDIA Tools Extension) ranges are CPU-side annotations pushed and popped by the host
process.  PyTorch's \texttt{torch.autograd.profiler.emit\_nvtx()} context manager inserts
\texttt{nvtxRangePushA(\allowbreak"aten::linear")} before and \texttt{nvtxRangePop()} after every
\texttt{aten::} dispatch call, labeling the host-side window during which that operator's CUDA
kernels are launched.  These annotations appear in the \nsys SQLite export as
\texttt{NVTX\_EVENTS} rows with CPU-domain timestamps.

For each CUDA stream observed in the \nsys trace, \sys
constructs a per-stream interval tree: a sorted list of NVTX ranges indexed by start time,
supporting $O(\log n)$ insertion and $O(k \log n)$ enclosure queries ($k$ = number of enclosing ranges).

\begin{algorithm}[t]
\caption{NVTX Enclosure Attribution (MEDIUM Confidence)}
\label{alg:nvtx}
\begin{algorithmic}[1]
\Require Kernel row $k$ with host-side launch timestamp $T_k$ and \texttt{host\_tid}
\Require \texttt{forest}: map from stream to per-stream interval tree
\Ensure Attribution at MEDIUM confidence, or fall through to heuristic
\State $\textit{ranges} \gets$ \texttt{forest}[$k.\textit{host\_tid}$].query\_enclosing($T_k$)
  \Comment{all NVTX ranges $[s,e]$ with $s \le T \le e$, outermost first}
\State $\textit{opRanges} \gets [\,]$
\For{$r \in \textit{reverse}(\textit{ranges})$}
  \Comment{walk innermost $\to$ outermost}
  \If{\texttt{is\_attributed\_op}($r.\textit{text}$)}
    \State append $r$ to $\textit{opRanges}$
      \Comment{\texttt{aten::}, \texttt{quantized::}, or custom torch.library op}
  \ElsIf{\texttt{"::"} \textbf{not in} $r.\textit{text}$}
    \State \textbf{break}
      \Comment{structural boundary (e.g.\ \texttt{ProfilerStep\#0}) --- stop the walk}
  \Else
    \State \textbf{continue}
      \Comment{namespaced but non-kernel (\texttt{prims::}, \texttt{torch::}) --- skip, keep walking}
  \EndIf
\EndFor
\If{$\textit{opRanges} = [\,]$}
  \State \textbf{fall through} to Inductor fusion enrichment (\secref{sec:heuristic}) or UNATTRIBUTED
\Else
  \State $k.\textit{operator} \gets \textit{opRanges}[0].\textit{text}$,
         \textit{confidence} $\gets$ MEDIUM
    \Comment{innermost attributed range}
  \State $k.\textit{is\_fused} \gets (|\textit{opRanges}| > 1)$;
         $k.\textit{all\_enclosing} \gets \textit{ranges}$
    \Comment{fused kernel spans multiple attributed operators}
\EndIf
\end{algorithmic}
\end{algorithm}

The host-side launch timestamp is used as the query point rather than the GPU execution timestamp,
because NVTX ranges are CPU-side events and the async gap between host launch
and GPU execution makes GPU timestamps unreliable for enclosure queries.
Kernels launched before the first NVTX range (JIT compilation artifacts during warm-up)
produce an empty enclosure query and fall through to UNATTRIBUTED or Inductor enrichment;
no special handling is required.  When a CUDA graph is replayed, new GPU timestamps are
generated with no corresponding NVTX events; \sys detects the absence of enclosing ranges
in the replay window and replays the attribution cache recorded during the graph capture phase.
When multiple NVTX ranges enclose a single kernel---indicating Triton fusion across operator
boundaries---Algorithm~\ref{alg:nvtx} sets \texttt{is\_fused=True} and stores all enclosing
ranges in \texttt{all\_enclosing\_ranges} for downstream analysis.

\subsection{Inductor Fusion Enrichment}
\label{sec:heuristic}

When NVTX enclosure fails (no enclosing range exists at the query point), \sys applies a
post-attribution enrichment pass that recovers the fused operator identity from Inductor
debug artifacts, rather than applying heuristic inference from kernel names alone.

\textbf{Inductor debug extraction.}  When \texttt{torch.compile(\allowbreak backend="inductor")} runs with
debug mode enabled, Inductor writes \texttt{output\_code.py} artifacts containing the generated
Triton kernel source.  Each kernel call site is preceded by a structured comment:

\begin{lstlisting}[style=pystyle]
# Topologically Sorted Source Nodes: [...],
# Original ATen: [aten.relu, aten.addmm]
def triton_poi_fused_relu_addmm_0(in_ptr0, ...):
    ...
\end{lstlisting}

\sys parses these structured comments to extract the list of constituent \texttt{aten::} operators,
producing a fusion map that contains the exact set of operators fused into each kernel.

\textbf{Enrichment as a post-attribution pass.}  This map is applied \emph{after} the NVTX
enclosure pass and never overrides a valid TORCH\_PROFILER or NVTX attribution:
\begin{enumerate}[leftmargin=1.5em]
  \item For kernels that remain UNATTRIBUTED after NVTX enclosure, a lookup in the fusion map
    upgrades them to Inductor-fusion attribution at MEDIUM confidence, with the
    source operator list populated from the \texttt{Original~ATen} comment.
  \item For already-attributed kernels (TORCH\_PROFILER or NVTX), the enrichment augments the
    fused operator list and marks the kernel as fused if multiple operators are present.
\end{enumerate}

Kernels matching neither NVTX enclosure nor the Inductor fusion map remain in
\texttt{unattributed\_\allowbreak kernels[]} with \texttt{attribution\_method = "unattributed"}.  For eager-mode
workloads or cuDNN-backed operators (\eg \texttt{nn.LSTM} via \texttt{aten::\_cudnn\_rnn}), which
produce neither NVTX brackets nor Inductor debug output, unattributed rates of 20--85\% are expected
and are surfaced explicitly in the profile output.

\subsection{Layer Deduplication}
\label{sec:dedup}

For models with repeated structure (\eg transformer decoder blocks), profiling every
layer independently costs $O(N)$ \ncu replay time, where $N$ is the total number of layer partitions.
\sys identifies structurally identical subgraphs by computing a canonical hash over each
partition's operator types and connectivity, grouping them into $K$ equivalence classes.
Only one representative per class is compiled and profiled; its hardware counter values
are propagated to all structural duplicates, reducing replay cost to $O(K)$.

Deduplication is transparent to the attribution pipeline: duplicate partitions are
tagged in the \nsys trace and skipped during \ncu replay, with metrics inherited from
the unique representative by positional index.  A correctness invariant is that
partition tags must be emitted only during the measured capture window and not during
warmup; otherwise warmup-phase kernels are misattributed to operators rather than
classified as pre-measurement initialization, artificially inflating attributed kernel
counts and biasing operator durations.

\subsection{Bottleneck Axes and Counter Selection}
\label{sec:metrics}

Collecting all 90+ \ncu counters incurs substantial overhead (multiple replay passes) and
produces profiles too large for operator-level analysis.  \sys selects 20 counters spanning
memory bandwidth, compute throughput, Tensor Core utilization, occupancy and latency hiding,
register pressure, and warp divergence, subject to two additional constraints:
cross-architecture availability on Ampere, Hopper, \emph{and} Blackwell (several counter
names were renamed in the Blackwell GB202 silicon), and non-redundancy (no two selected
counters correlate perfectly).  The full counter list with bottleneck axis, and aggregation
method is in \tabref{tab:counters}
(\secref{sec:appendix:counters}).  Seven counters require name fallbacks; \sys queries all
variants and uses the first that returns a non-null value.

\subsubsection{Duration-Weighted Aggregation}
\label{sec:aggregation}

An operator may dispatch multiple kernels with heterogeneous durations.  For rate and utilization
metrics (rows marked DW-mean in \tabref{tab:counters}), a naive arithmetic mean would give a
short-lived auxiliary kernel the same weight as the long-running compute kernel.  \sys uses
duration-weighted aggregation instead:
\begin{equation}
  \overline{m}_\text{op} = \frac{\displaystyle\sum_{i} m_i \cdot d_i}{\displaystyle\sum_{i} d_i}
  \label{eq:dwmean}
\end{equation}
where $m_i$ and $d_i$ are the metric value and duration of the $i$-th kernel attributed to the
operator.  This ensures that the dominant kernel---the primary optimization target---contributes
proportionally to the operator-level summary.  For additive quantities (bytes, instruction counts,
spills), \sys uses plain summation.  For per-kernel constants (register count, shared memory),
\sys takes the maximum across all kernels, since the highest-pressure kernel limits occupancy.

\section{Evaluation: Attribution Quality and Profile-Guided Optimization}
\label{sec:experiments}

We evaluate \sys on two axes: (1) \emph{coverage}---what fraction of kernel runtime is
correctly attributed to \texttt{aten::} operators at each confidence tier; and
(2) \emph{utility}---whether attributed profiles provide sufficient signal to ground and
verify optimization decisions.  For the utility axis, we implement a profile-guided
optimization workflow in which per-operator counter values from \texttt{profile.json} drive
the selection of FX graph transformations; the optimization layer is one downstream
application of the attribution pipeline---others include hardware-aware neural architecture
search, performance regression detection in CI, and kernel-level profiling of serving stacks
such as vLLM~\cite{vllm2023} or SGLang~\cite{sglang2024}.

\subsection{Experimental Setup}

\textbf{Hardware.}  All experiments run on a single NVIDIA RTX PRO 6000 Blackwell GPU; full hardware specifications are in \secref{sec:appendix:hwspec}.

\textbf{Workloads.}  \tabref{tab:workloads} describes the three workloads evaluated.  Each
isolates a distinct operator category and hardware challenge; batch sizes reflect
the hardcoded defaults in each workload file and were not adjusted for this evaluation.

\begin{table}[h]
\centering
\caption{Case study workloads.}
\label{tab:workloads}
\small
\begin{tabular}{@{}lcp{5.5cm}@{}}
\toprule
\textbf{Workload} & \textbf{Batch} & \textbf{Dimensions} \\
\midrule
GPT-2 (12-layer)   & 4  & hidden=768, FFN=3072, 12 heads, seq=128 \\
SDPA Attention     & 8  & seq=512, dim=512, 8 heads \\
LSTM Seq.\ Encoder & 32 & seq=128, hidden=512, 2 layers \\
\bottomrule
\end{tabular}
\end{table}

\textbf{Measurement protocol.}  Reported duration fields are \nsys CUPTI GPU-domain
kernel times.  Operator summaries sum the \nsys durations of kernels attributed to that
operator; workload speedups sum profiled forward-pass kernel time from the same capture.
\ncu replay supplies hardware counters only, including the raw
\texttt{gpu\_\_time\_duration.sum} counter, and replay timing is excluded from speedup
reporting.  Each workload runs 2 warm-up iterations followed by 2 measured iterations;
results are point estimates from those measured iterations.

\textbf{Speedup scope.}  Speedup figures are computed over profiled forward-pass
kernel wall time at locked clocks.  Operator-level diagnosis uses attributed operator
records, while unattributed kernels are reported separately and are not silently dropped.
The LSTM Inductor baseline is non-representative: \texttt{torch.compile} triggers a hard
Dynamo graph break on \texttt{nn.LSTM}, decomposing it into 1{,}280 small per-timestep FP32
GEMMs rather than a fused cuDNN RNN call; the result reflects correcting this structural
mismatch via cuDNN re-dispatch (\secref{sec:cs:lstm}), not a conventional FX graph
optimization.

\textbf{Baselines.}  Each workload is compared as:
\begin{enumerate*}[label=(\arabic*)]
  \item \emph{Inductor FP32 baseline}: \texttt{torch.compile(\allowbreak backend="inductor")} with
    default settings, FP32 precision;
  \item \emph{Optimized}: workload-specific backend applying BF16 promotion and other
    hardware-evidence-motivated patterns.
\end{enumerate*}
Speedups are reported as \emph{Inductor FP32 baseline} $\div$ \emph{Optimized} in attributed
kernel wall time at locked clocks.
Each optimized backend applies passes as a suite rather than in isolation because the
passes target orthogonal hardware bottleneck axes---dtype precision (BF16 promotion),
memory access pattern (SDPA canonicalization), and constant propagation (weight
freezing)---and are selected jointly by the attributed counter evidence.
Measuring the combined effect is the relevant outcome for the profile-guided workflow;
the per-pass counter deltas reported in each case study (\secref{sec:cs:gpt2}--\secref{sec:cs:lstm})
serve as per-pass verification that each transformation eliminated its target bottleneck.

\subsection{Attribution Coverage}
\label{sec:coverage}

\begin{figure}[H]
\centering
\begin{tikzpicture}
\begin{axis}[
  xbar stacked,
  width=0.88\columnwidth,
  height=4.0cm,
  bar width=10pt,
  ymin=-0.5,
  ymax=2.5,
  xmin=0, xmax=100,
  xlabel={Kernel runtime attributed (\%)},
  xlabel style={font=\small},
  ytick=data,
  yticklabels={
    LSTM Seq.\ Enc.,
    SDPA Attention,
    GPT-2
  },
  yticklabel style={font=\footnotesize},
  tick label style={font=\footnotesize},
  axis x line=bottom,
  axis y line=left,
  xmajorgrids=true,
  grid style={gray!20},
  legend style={
    at={(0.5,-0.48)}, anchor=north,
    font=\scriptsize, draw=gray!50,
    fill=white, fill opacity=0.9,
    legend columns=3,
  },
]

\addplot[fill=highcol, draw=highcol!80!black] coordinates {
  (4,   0)   
  (35,  1)   
  (40,  2)   
};

\addplot[fill=medcol!70, draw=medcol!80!black] coordinates {
  (8,   0)   
  (62,  1)   
  (57,  2)   
};

\addplot[fill=ucol!80, draw=gray!60] coordinates {
  (88,  0)   
  (3,   1)   
  (3,   2)   
};

\legend{HIGH (torch.profiler IOM), MEDIUM (NVTX / Inductor), UNATTRIBUTED}

\end{axis}
\end{tikzpicture}
\caption{\textbf{Attribution coverage for the three case study workloads.}  GPT-2 and SDPA
  Attention achieve $>$95\% attributed runtime with the correlation pass active.  LSTM's
  $>$85\% UNATTRIBUTED fraction correctly reflects opaque cuDNN RNN dispatch
  (see \secref{sec:cs:lstm}).  Inductor fusion enrichment (\secref{sec:heuristic})
  typically adds 5--15 percentage points for compiled workloads.}
\label{fig:attribution}
\end{figure}
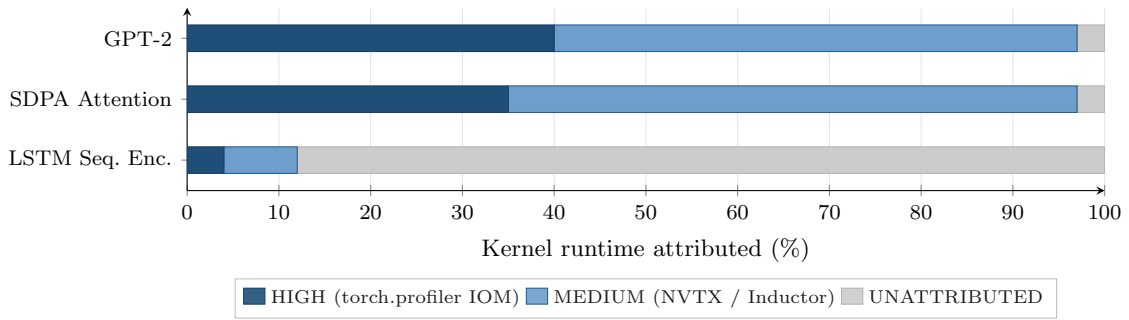

Attribution coverage for the three case study workloads is shown in \figref{fig:attribution}.
GPT-2 and SDPA Attention achieve near-complete attribution with the correlation pass active.
LSTM's $>$85\% unattributed fraction is the expected result for that dispatch path; the cause
and corrective strategy are examined in \secref{sec:cs:lstm}.

\subsection{Optimization Infrastructure}
\label{sec:threelevel}

The \texttt{torch.compile()} pipeline lowers a model through successive IR stages---Dynamo
capture, AOTAutograd decomposition, and Inductor code generation---each of which exposes
a different graph representation.  Profile-guided optimizations are applied at one of three
corresponding levels, depending on which representation exposes the structure required by
each transformation.
\textbf{Level~1} (pre-decomposition) targets high-level operator patterns such as
multi-head attention or QKV fusion that are only visible before AOTAutograd lowers
shared activation nodes into per-consumer copies.  \textbf{Level~2} (Aten IR)
targets fully decomposed operator graphs and is suited for dtype casts, weight
folding, and memory-layout annotation.  \textbf{Level~3} (Inductor configuration)
handles non-graph directives such as autotuning and weight freezing that are
expressed as compiler flags rather than graph rewrites.  Passes at each level
compose without interfering with those at other levels.

\subsection{Optimization Case Studies}
\label{sec:validation}

The following case studies demonstrate what per-operator attributed profiles enable that
traditional tools cannot: \texttt{torch.profiler} provides operator timing without hardware
counters; \nsys provides kernel timing without operator attribution; \ncu provides hardware
counters without operator attribution.  Each case study identifies the specific counter
symptom read from \sys's per-operator attributed profile, the transformation it motivates,
and the counter delta that confirms bottleneck elimination.  \tabref{tab:speedup} summarizes
profiled-kernel-time speedups at locked clocks.

\begin{table}[h]
\centering
\caption{Forward-pass profiled kernel wall time and speedup on NVIDIA RTX PRO 6000 Blackwell.
  All times are sums of \nsys CUPTI kernel durations at locked GPU clocks (not \ncu replay).
  Speedup = Inductor FP32 $\div$ Optimized.  Each result is a point estimate from
  2 measured iterations.}
\label{tab:speedup}
\small
\setlength{\tabcolsep}{5pt}
\begin{tabular}{lrrr}
\toprule
\textbf{Workload} & \textbf{Inductor FP32 Baseline} & \textbf{Optimized}
                  & \textbf{Speedup} \\
\midrule
GPT-2 (12-layer)        & 7,313\,$\mu$s  & 4,161\,$\mu$s  & \textbf{1.76$\times$} \\
SDPA Attention          & 906\,$\mu$s    & 405\,$\mu$s    & \textbf{2.24$\times$} \\
LSTM Seq.\ Encoder$^\dagger$ & 14,531\,$\mu$s & 4,046\,$\mu$s  & \textbf{3.59$\times$} \\
\bottomrule
\end{tabular}
\vspace{4pt}

\noindent{\footnotesize $^\dagger$~Inductor baseline is non-representative: \texttt{nn.LSTM} triggers a Dynamo graph break, producing 1{,}280 small per-timestep GEMMs instead of a fused cuDNN RNN call. The LSTM case is diagnostic rather than a conventional FX optimization: restricting to attributed operator records gives 3.63$\times$, while including the optimized-side unattributed kernel tail gives the 3.59$\times$ profiled-kernel speedup shown here.}
\end{table}

\subsection{Case Study 1 --- GPT-2: Tensor Core Idle $\to$ BF16 Promotion}
\label{sec:cs:gpt2}

A workload-level \ncu report exposes \texttt{tensor\_core\_active\_pct = 0\%} as a global
aggregate but cannot identify which operator family holds that idle compute or what fraction
of total runtime it represents.  \tabref{tab:transformer} shows the per-operator attributed
breakdown: 87.5\% of attributed runtime (6{,}401 of 7{,}313\,$\mu$s) resides in the GEMM
family alone, with \texttt{tensor\_core\_active\_pct = 0\%} and 210 registers/thread.  This
localization makes the optimization target and expected gain calculable before writing any
backend code.

\begin{table}[h]
\centering
\caption{Per-operator-class attributed time for GPT-2 (RTX PRO 6000 Blackwell, B=4, seq=128,
  locked clocks).  Baseline: Inductor FP32 (built-in dedup backend).  Optimized: BF16 dtype
  promotion + SDPA canonicalization + Inductor freezing (\texttt{gpt2\_opt} backend).
  ``Triton casts'' are fp32$\leftrightarrow$bf16 cast kernels introduced by the bf16 promotion
  that the cast-cancellation pass (OPT-4) was unable to eliminate.}
\label{tab:transformer}
\small
\begin{tabular*}{\textwidth}{@{\extracolsep{\fill}}lrrr}
\toprule
\textbf{Operator class} & \textbf{Baseline} & \textbf{Optimized} & \textbf{Speedup} \\
\midrule
GEMM family (\texttt{cutlass\_80\_simt\_sgemm}, 96 kernels) & 6,401\,$\mu$s & 2,695\,$\mu$s & \textbf{2.37$\times$} \\
Attention (\texttt{fmha\_cutlassF\_f32})                    &   485\,$\mu$s &   408\,$\mu$s & 1.19$\times$ \\
LayerNorm                                                   &   149\,$\mu$s &   120\,$\mu$s & 1.24$\times$ \\
Triton casts + elementwise + epilogues                      &   279\,$\mu$s &   938\,$\mu$s & new overhead \\
\midrule
\textbf{Total}                                              & \textbf{7,313\,$\mu$s} & \textbf{4,161\,$\mu$s} & \textbf{1.76$\times$} \\
\bottomrule
\end{tabular*}
\end{table}

\subsubsection*{Diagnosis}

BF16 dtype promotion rerouted all 96 GEMM kernels from the FP32 SIMT path to the BF16
Tensor Core path; \tabref{tab:gpt2_counters} confirms the counter shift.  The attention
speedup derives from SDPA canonicalization removing the 48-launch causal-mask subgraph,
which is only visible as a distinct operator class at the functional FX level before
AOTAutograd decomposition (\secref{sec:threelevel}).  The residual
fp32$\leftrightarrow$bf16 cast overhead (+0.66\,ms, 288 new Triton kernels) is directly
auditable from the ``Triton casts'' row in \tabref{tab:transformer}---the attributed
profile identifies not only the gain but the exact overhead introduced by the optimization.

\begin{table}[h]
\centering
\caption{Key hardware counter deltas for GPT-2 GEMM family (RTX PRO 6000 Blackwell).
  GEMM counters are duration-weighted aggregates (\equref{eq:dwmean}).}
\label{tab:gpt2_counters}
\small
\setlength{\tabcolsep}{4pt}
\begin{tabular}{lrrl}
\toprule
\textbf{Counter} & \textbf{Baseline} & \textbf{Optimized} & \textbf{Bottleneck Resolved} \\
\midrule
TC active, GEMMs (\%)          & 0.0\%        & 40.5\%       & FP32 SIMT $\to$ BF16 Tensor Core \\
Registers/thread (GEMMs)       & 210          & 96           & BF16 tiles need fewer registers \\
Achieved occupancy, GEMMs (\%) & 16.5\%       & ${\sim}$30\% & Register pressure relief \\
SM throughput, GEMMs (\%)      & ${\sim}$40\% & ${\sim}$65\% & Correct compute pipeline \\
\bottomrule
\end{tabular}
\end{table}

\subsection{Case Study 2 --- SDPA Attention: Low Occupancy and QKV Fusion}
\label{sec:cs:sdpa}

A workload-level occupancy reading reveals low average occupancy but cannot identify which
operator is responsible or why.  The attributed profile localizes occupancy\,=\,16.6\% to
the Q/K/V projection GEMMs specifically, where the baseline concentrates 96.7\% of attributed
runtime: eight \texttt{aten::mm} GEMMs (66.2\%, TC-idle) and two SDPA calls (30.5\%).
Crucially, the FX graph shows all three projections read from a single shared
\texttt{LayerNorm} activation---the structural precondition for QKV fusion.  This causal chain
(per-operator counter $\to$ graph-structural diagnosis $\to$ specific FX pass) is not available
from aggregate profiling alone.

Three passes were applied.  \textbf{BF16 dtype promotion} rerouted the GEMMs off the FP32
SIMT path and switched the SDPA kernel from \texttt{fmha\_cutlassF\_f32} to
\texttt{fmha\_cutlassF\_bf16}.  \textbf{QKV fusion} merged the three projections into a
single \texttt{[512, 1536]} GEMM (three serial 128-block launches $\to$ one 384-block launch),
raising achieved occupancy from 16.6\% to 73.9\%.  \textbf{Inductor freezing} arranged
weights in aligned memory order for bf16 GEMM autotuning.  Newly-introduced auxiliary kernels
account for 9.1\% of the optimized profile runtime, bounding the observed 2.24$\times$
speedup.

\begin{table}[h]
\centering
\caption{Key hardware counter deltas for SDPA Attention (RTX PRO 6000 Blackwell).
  GEMM counters are duration-weighted aggregates (\equref{eq:dwmean}).}
\label{tab:sdpa_counters}
\small
\setlength{\tabcolsep}{4pt}
\begin{tabular}{lrrl}
\toprule
\textbf{Counter} & \textbf{Baseline} & \textbf{Optimized} & \textbf{Bottleneck Resolved} \\
\midrule
TC active, QKV GEMMs (\%)  & 0.0\%  & 19--21\% & FP32 SIMT $\to$ BF16 Tensor Core \\
Occupancy, QKV GEMMs (\%)  & 16.6\% & 73.9\%   & 3 serial $\to$ 1 wide launch (QKV fusion) \\
Attention kernel            & \texttt{fmha\_f32} & \texttt{fmha\_bf16} & BF16 promotes to bf16 fmha path \\
\bottomrule
\end{tabular}
\end{table}

\subsection{Case Study 3 --- LSTM Sequence Encoder: Attribution Tier as Structural Diagnostic}
\label{sec:cs:lstm}

A kernel-level timer would expose 1{,}280 slow small-matrix GEMMs (M=32, TC-idle) and
motivate direct kernel optimization---tiling, BF16 casts, epilogue fusion.  The attribution
tier changes the diagnosis: $>$85\% UNATTRIBUTED from a cuDNN-opaque dispatch path identifies
the LSTM body as unreachable by any FX-level transformation.  This structural signal is not
available from timing-only profiling.

Under \texttt{torch.compile(\allowbreak backend="inductor")}, \texttt{nn.LSTM} triggers a hard Dynamo
graph break (\texttt{graph\_count = 0}) via \texttt{torch.\_VF.lstm}, a C extension entry
point Dynamo cannot graph-capture.  Inductor decomposes the LSTM into a per-timestep unrolled
loop: 1,280 small-matrix FP32 GEMMs (M=32, \texttt{tensor\_core\_active\_pct = 0.0},
achieved occupancy 9.4\%) with 1,280 matching \texttt{cublasLt::\allowbreak splitKreduce\_kernel}
epilogues.  Attribution is $>$85\% UNATTRIBUTED because cuDNN's internal RNN kernels carry no
NVTX annotations.

The corrective strategy is structural: the custom backend routes the recurrent region to
cuDNN's fused Tensor-Core RNN, which batches the input-to-hidden projection across all 128
timesteps in one launch (M=4096 instead of M=32), eliminating all 1,280
\texttt{splitKreduce} epilogues and reducing attributed execution time by 2.26\,ms.  All FX
passes report \texttt{NOT\_APPLIED}---the backend callback is never invoked for the
eagerly-dispatched LSTM region---yet this is the largest attributed-region speedup in the
evaluation suite (3.63$\times$), demonstrating that the attributed pipeline correctly routes
structural problems to structural fixes.

\begin{table}[h]
\centering
\caption{Key hardware counter deltas for LSTM Sequence Encoder (RTX PRO 6000 Blackwell).
  GEMM counters are duration-weighted aggregates (\equref{eq:dwmean}).}
\label{tab:lstm_counters}
\small
\setlength{\tabcolsep}{4pt}
\begin{tabular}{lrrl}
\toprule
\textbf{Counter} & \textbf{Baseline} & \textbf{Optimized} & \textbf{Bottleneck Resolved} \\
\midrule
TC active, recurrent GEMMs (\%)  & 0.0\%   & 23.8\% & Scalar SIMT $\to$ cuDNN fused RNN \\
Occupancy, recurrent GEMMs (\%)  & 9.4\%   & 39.4\% & M=32 $\to$ M=4096 (batch timesteps) \\
\texttt{splitKreduce} epilogues  & 1{,}280 & 0      & Eliminated by cuDNN routing \\
\bottomrule
\end{tabular}
\end{table}

\section{Conclusion}
\label{sec:conclusion}

The attribution gap between framework profilers and GPU hardware counter tools exists not
because the relevant data is absent, but because no stable join key spans the two collection
passes: kernel names are non-unique identifiers, timestamps are incompatible across tool clock
domains, and manual bridging costs $O(\text{operators})$ effort per model variant.
\sys resolves all three failures through three complementary attribution paths---CUPTI
External Activity correlation (HIGH confidence, causal dispatch linkage), NVTX temporal
enclosure via per-stream interval trees (MEDIUM), and Inductor debug-artifact enrichment
(MEDIUM)---with invocation-order matching joining \ncu hardware counters to \nsys kernel
records by execution order rather than timestamp, bypassing clock-domain incompatibility
without any cross-tool timestamp comparison.
GPU clock locking eliminates the systematic speedup bias that arises when a lower-power
optimized workload sustains a higher boost frequency than the baseline; reported duration
ratios are clock-immune by construction.
A curated 20-counter metric set with duration-weighted aggregation and
architecture-specific fallbacks covers all hardware bottleneck axes, and layer deduplication
reduces \ncu replay time by a factor of $N/K$ for models with $N$ repeated layers across $K$
unique structural classes ($12{\times}$ for GPT-2's 12 identical transformer blocks).

To establish that attributed profiles carry sufficient per-operator counter signal to
identify and verify the elimination of specific hardware bottlenecks, we applied \sys as
the upstream profiling component in a profile-guided FX optimization workflow---one
downstream application among others including hardware-aware neural architecture search,
serving-stack kernel profiling, and CI performance regression detection.
Across three forward-pass case studies on a single NVIDIA RTX PRO 6000 Blackwell GPU
(point estimates from 2 measured iterations at locked clocks), each optimization decision
was grounded in a specific counter reading from the attributed profile and each claimed
improvement was verified by re-profiling at identical locked clocks.  The two compiled
optimization case studies show profiled-kernel-time speedups of
\textbf{1.76$\times$} on GPT-2 (\texttt{tensor\_core\_active\_pct}: 0\%$\to$40.5\%)
and \textbf{2.24$\times$} on SDPA Attention (occupancy 16.6\%$\to$73.9\%).
The LSTM case is diagnostic rather than a conventional FX optimization: the $>$85\%
UNATTRIBUTED attribution tier identifies the LSTM body as structurally unreachable by
FX-level transformation and routes the diagnosis to cuDNN re-dispatch.  That structural
fix yields a \textbf{3.63$\times$} attributed-region speedup, or \textbf{3.59$\times$}
when including the optimized-side unattributed kernel tail; the corrective action is
not taken by \sys and should not be interpreted as an end-to-end FX-optimization result.

\paragraph{Limitations.}
The evaluation covers three workloads on a single GPU architecture; no multi-GPU
configurations were evaluated, and generalizability across architectures and workload
scales remains unverified.
All speedup figures are point estimates from 2 measured iterations over profiled kernel
time; run-to-run variance and end-to-end wall-clock speedups are not characterized.
The pipeline surfaces per-operator counter evidence but does not prescribe optimization
actions; the Claude Code plugin (\appref{sec:appendix:plugin}) automates the closed-loop
protocol at the prototype level, though whether agent-generated backends reproduce the
manually-derived speedups reported here is left for future work.
Attribution coverage degrades for cuDNN-backed operators (\texttt{nn.LSTM},
\texttt{nn.\allowbreak MultiheadAttention} under cuDNN) where internal kernel names carry no
NVTX correspondence and no Inductor debug metadata is available; expected unattributed rates
of 80--90\% make hardware-counter-driven optimization inapplicable at the FX level, and
bottleneck diagnosis must fall back to kernel-name inference.
Invocation-order matching requires a deterministic workload: models with conditional control
flow or dynamic dispatch can violate the \texttt{(kernel\_name, invocation\_index)} invariant
without producing a shape error; \sys validates invocation counts before issuing counter
assignments and raises \texttt{InvocationCountMismatchError} on violation.
\ncu serializes kernels for replay, making concurrent multi-stream profiles require separate
per-stream runs.

\paragraph{Future work.}
Multi-GPU attribution with NCCL stream synchronization events would extend coverage to
distributed workloads, and online profiling via CUPTI Perfworks streaming would eliminate
the separate \ncu replay pass, enabling attribution during any workload without a
dedicated replay run.
Cross-architecture validation on Ampere and Hopper is a near-term priority: counter name
fallbacks are implemented but attribution quality has not been verified at evaluation scale
on those architectures.
Counter-attributed profiles also provide a precise regression signal for CI-integrated
performance testing: a hardware-counter shift on a code change---rather than a latency
shift alone---distinguishes architectural regressions from scheduling noise, a downstream
application of the attribution output that does not require any optimization framework.
Downstream of the attribution layer, automatic bottleneck-to-transformation mapping---
inferring which graph passes are warranted from a counter symptom without expert
interpretation---would close the loop to a fully autonomous optimization pipeline; a
dataset of (workload, counter reading, transformation, measured outcome) tuples would
make this a concrete supervised learning problem.
Finally, generalizing the counter name mapping and pass library to AMD ROCm
(\texttt{rocprof} + \texttt{rocm-smi}) would extend \sys beyond NVIDIA hardware.

\newpage
\bibliographystyle{plainnat}
\bibliography{reference}
\newpage
\appendix
\onecolumn
\section{Reproduction and Additional Details}
\label{sec:appendix:exp}

\subsection{Hardware Specification}
\label{sec:appendix:hwspec}

All experiments run on a single GPU:
\begin{itemize}[leftmargin=1.5em]
  \item \textbf{NVIDIA RTX PRO 6000 Blackwell}: ${\sim}$188 SMs (GB202),
    96\,GB GDDR7 (${\sim}$1.8\,TB/s memory bandwidth),
    5th-generation Tensor Cores (HMMA BF16/FP8), Blackwell architecture (sm\_120).
    Note: the \texttt{warp\_cycles\_per\_instruction} counter is absent from the
    Blackwell counter set; \texttt{eligible\_\allowbreak cycles\_pct} serves as the
    latency-bound indicator instead.
\end{itemize}

Software requirements: NVIDIA GPU (Ampere minimum), \nsys $\ge$ 2024.6,
\ncu $\ge$ 2025.4.1, PyTorch 2.11+, and CUDA 12.8.
Reproduction scripts, \texttt{profile.json} pairs, and per-example counter comparisons are provided in the accompanying repository here: 
\url{https://github.com/yottalabsai/Profiler}

\subsection{Hardware Counter Reference}
\label{sec:appendix:counters}

\begin{table}[h]
\centering
\caption{The 20 hardware counters collected by \sys, their bottleneck axis, and
what they measure.  ``Agg.'' denotes the aggregation method
  applied when multiple kernels are attributed to one operator (see \secref{sec:aggregation}).
  $\dagger$~\texttt{warp\_cycles\_per\_instr} is absent from the Blackwell counter set (sm\_120);
  \texttt{eligible\_\allowbreak cycles\_pct} (counter~14) serves as the latency-bound indicator on the
  RTX PRO 6000 used for all experiments reported here.}
\label{tab:counters}
\scriptsize
\setlength{\tabcolsep}{4pt}
\begin{tabular}{clllc}
\toprule
\textbf{\#} & \textbf{Profile Field} & \textbf{Bottleneck Axis} & \textbf{Measures} & \textbf{Agg.}\\
\midrule
1  & \texttt{gpu\_time\_duration\_ns}      & Latency          & GPU wall time                & SUM \\
2  & \texttt{dram\_bytes\_read}            & Memory BW        & DRAM read bytes              & SUM \\
3  & \texttt{dram\_bytes\_written}         & Memory BW        & DRAM write bytes             & SUM \\
4  & \texttt{memory\_throughput\_pct}      & Memory BW        & Memory subsystem \% peak     & DW-mean \\
5  & \texttt{dram\_throughput\_pct}        & Memory BW        & DRAM \% of peak BW           & DW-mean \\
6  & \texttt{mem\_busy\_pct}              & L1 pipeline      & L1 pipeline \% peak active   & DW-mean \\
7  & \texttt{l1\_hit\_rate}               & Cache            & L1 cache hit rate (\%)       & DW-mean \\
8  & \texttt{l2\_hit\_rate}               & Cache            & L2 cache hit rate (\%)       & DW-mean \\
9  & \texttt{sm\_throughput\_pct}         & Compute          & SM \% of peak throughput     & DW-mean \\
10 & \texttt{tensor\_core\_active\_pct}   & Tensor Cores     & TC pipeline active (\%)      & DW-mean \\
11 & \texttt{sm\_active\_cycles}          & Compute          & Total SM active cycles       & SUM \\
12 & \texttt{achieved\_occupancy}         & Occupancy        & Warps active \% of peak      & DW-mean \\
13$\dagger$ & \texttt{warp\_cycles\_per\_instr}    & Latency          & Cycles per issued instruction & DW-mean \\
14 & \texttt{eligible\_\allowbreak cycles\_pct}       & Latency          & Eligible warp issue rate (\%) & DW-mean \\
15 & \texttt{executed\_instructions}      & Throughput       & Total executed instructions  & SUM \\
16 & \texttt{ipc\_active}                 & Throughput       & IPC when SM active           & DW-mean \\
17 & \texttt{avg\_threads\_per\_warp}     & Divergence       & Avg threads active per warp  & DW-mean \\
18 & \texttt{registers\_per\_thread}      & Reg.\ pressure   & Registers allocated / thread & MAX \\
19 & \texttt{local\_memory\_spills}       & Reg.\ spills     & Spill events to L1 local mem & SUM \\
20 & \texttt{dynamic\_smem\_per\_block}   & SMEM pressure    & Dynamic shared mem / block   & MAX \\
\bottomrule
\end{tabular}
\end{table}

\subsection{Prototype Automation via Claude Code Plugin}
\label{sec:appendix:plugin}

As a prototype implementation, \sys includes a Claude Code plugin that automates the
closed-loop protocol via a single \texttt{/optimize workload.py} command.  The plugin orchestrates five
specialized LLM agents---\textbf{capture-agent}, \textbf{optimization-strategist},
\textbf{backend-engineer}, \textbf{validation-agent}, and \textbf{/report}---covering all
pipeline stages from baseline capture through cross-profile comparison.

\subsection{Annotated \texttt{profile.json} Schema}
\label{sec:appendix:schema}

\texttt{profile.json} is the central output of \sys: a JSON document that maps each
dispatched PyTorch operator to its constituent CUDA kernels, per-kernel \nsys durations,
and aggregated hardware metrics.  \tabref{tab:schema_fields} describes every top-level field; the listing below
shows a representative single-operator excerpt from the GPT-2 baseline
(\texttt{aten::mm}, 512$\times$768 $\times$ 768$\times$768 weight projection, NVTX-attributed).
The \texttt{aggregated} block condenses per-kernel counter values into a single
operator-level record using the duration-weighted mean (\equref{eq:dwmean}); these are the
values consumed by downstream analysis tools and the Claude Code plugin.
The \texttt{duration\_ns} and \texttt{aggregated.total\_duration\_ns} fields are
\nsys-derived timing fields; \ncu metrics, including
\texttt{gpu\_\_time\_duration.sum}, are stored only inside \texttt{metrics.raw}.

\begin{lstlisting}[basicstyle=\scriptsize\ttfamily, breaklines=true, frame=single,
                   xleftmargin=2pt, xrightmargin=2pt]
{
  "schema_version": "1.0",
  "capture_metadata": {
    "model_name": "gpt2",
    "device_name": "NVIDIA RTX PRO 6000 Blackwell",
    "compile_mode": "inductor",
    "capture_timestamp_utc": "2026-06-01T00:53:25Z"
  },
  "operators": [
    {
      "operator_id": "aten::mm_26",
      "operator_name": "aten::mm",
      "call_index": 26,
      "is_fused": false,
      "fused_with": [],
      "kernels": [
        {
          "kernel_id": "k_00880",
          "kernel_name": "cutlass_80_simt_sgemm_128x32_8x5_nn_align1",
          "attribution_method": "nvtx",
          "confidence": "medium",
          "duration_ns": 32736,
          "stream_id": 7,
          "grid_dim": [128, 1, 5],
          "block_dim": [128, 1, 1],
          "metrics": { "...": "20 raw ncu counter fields (architecture-specific names)" }
        }
      ],
      "aggregated": {
        "total_duration_ns": 32736,
        "kernel_count": 1,
        "tensor_core_active_pct": 0.0,
        "achieved_occupancy": 18.26,
        "sm_throughput_pct": 37.6,
        "dram_throughput_pct": 7.16,
        "l2_hit_rate": 89.55,
        "registers_per_thread": 80.0,
        "eligible_cycles_pct": 40.6,
        "local_memory_spills": 0,
        "warp_cycles_per_instruction": null
      }
    }
  ],
  "unattributed_kernels": []
}
\end{lstlisting}

Key fields: \texttt{attribution\_method} records the tier that assigned operator identity
(\texttt{"nvtx"}, \texttt{"inductor\_fusion"}, or \texttt{"torch\_profiler\_correlation"});
\texttt{confidence} mirrors the tier level (\texttt{"high"} or \texttt{"medium"});
\texttt{is\_fused} and \texttt{fused\_with} list all \texttt{aten::} ops fused into the
same Triton kernel.  The \texttt{null} value for \texttt{warp\_cycles\_per\_instruction}
reflects the absent Blackwell counter.
The \texttt{metrics.raw} sub-object stores all architecture-specific ncu column names
exactly as emitted by \texttt{ncu --import --csv}; the \texttt{aggregated} block outlines an operator-level metric summary of the attributed kernels.

\begin{table}[h]
\centering
\caption{Top-level \texttt{profile.json} field reference.}
\label{tab:schema_fields}
\scriptsize
\setlength{\tabcolsep}{4pt}
\begin{tabular}{lp{3.5cm}p{7.5cm}}
\toprule
\textbf{Field} & \textbf{Type} & \textbf{Meaning} \\
\midrule
\texttt{schema\_version}       & string        & Schema revision for forward-compatibility checks \\
\texttt{capture\_metadata}     & object        & Device name, torch/CUDA versions, nsys/ncu report paths \\
\texttt{operators[]}           & array         & One entry per attributed operator dispatch \\
\quad\texttt{operator\_id}     & string        & Unique key: \texttt{aten::op\_N} \\
\quad\texttt{operator\_name}   & string        & Canonical \texttt{aten::} name (with size annotations) \\
\quad\texttt{call\_index}      & int           & Dispatch ordinal within the measured window \\
\quad\texttt{is\_fused}        & bool          & True if kernel spans $>$1 aten op (Inductor fusion) \\
\quad\texttt{fused\_with}      & string[]      & Co-fused \texttt{aten::} op names \\
\quad\texttt{kernels[]}        & array         & Per-kernel records with raw ncu metrics \\
\quad\texttt{aggregated}       & object        & Duration-weighted operator-level metric summary \\
\texttt{unattributed\_kernels[]} & array       & Kernels matching no attribution path \\
\bottomrule
\end{tabular}
\end{table}

\subsection{Supplementary Figures}
\label{sec:appendix:candidates}

\begin{figure}[H]
\centering
\begin{tikzpicture}[font=\small, >=Stealth,
  stage/.style={draw, rounded corners=4pt, fill=white,
                minimum width=3.6cm, minimum height=0.75cm,
                align=center, inner sep=5pt, font=\small},
  hook/.style={draw, rounded corners=3pt, align=left,
               inner sep=5pt, font=\scriptsize},
  arr/.style={-Stealth,thick},
  harr/.style={-Stealth, thick, dashed},
]

  \node[stage, fill=gray!10] (in)     at (0,0)    {\texttt{workload.py}};
  \node[stage, fill=blue!8]  (dynamo) at (0,-2.2)  {Dynamo FX\\graph capture};
  \node[stage, fill=blue!8]  (aot)    at (0,-4.4)  {AOTAutograd\\decomposition};
  \node[stage, fill=blue!8]  (ind)    at (0,-6.6)  {Inductor\\code generation};
  \node[stage, fill=gray!10] (out)    at (0,-8.8)  {Triton / CUDA kernels};

  \draw[arr] (in)     -- (dynamo);
  \draw[arr] (dynamo) -- (aot);
  \draw[arr] (aot)    -- (ind);
  \draw[arr] (ind)    -- (out);

  \node[hook, fill=red!8, draw=red!50] (l1) at (5.5,-2.2)
    {\textbf{Level 1} (pre-decomposition)\\
     \textbullet\ QKV fusion (\textit{SDPA case study})\\
     \textbullet\ Attention canonicalization};
  \draw[harr, red!60] (l1.west) -- (dynamo.east);

  \node[hook, fill=orange!8, draw=orange!50] (l2) at (5.5,-4.4)
    {\textbf{Level 2} (Aten IR)\\
     \textbullet\ BF16 dtype promotion (\textit{all case studies})\\
     \textbullet\ Weight folding, layout annotation};
  \draw[harr, orange!60] (l2.west) -- (aot.east);

  \node[hook, fill=green!8, draw=green!50!black] (l3) at (5.5,-6.6)
    {\textbf{Level 3} (Inductor config)\\
     \textbullet\ Weight freezing (\textit{all case studies})\\
     \textbullet\ Autotuning mode selection};
  \draw[harr, green!60!black] (l3.west) -- (ind.east);

  \node[font=\scriptsize, gray, align=center] at (0,-9.6)
    {Passes at each level compose\\without interfering across levels};

\end{tikzpicture}
\caption{\textbf{Three-level FX pass hook points.}
  The \texttt{torch.compile()} pipeline lowers a model through three IR stages.
  Level~1 passes must run before AOTAutograd because structural patterns such as
  a shared \texttt{LayerNorm} output feeding Q, K, and V projections---the
  precondition for QKV fusion---are decomposed into per-consumer copies by
  AOTAutograd and become invisible afterward.  Level~2 (Aten IR) applies
  dtype and layout transforms on the fully decomposed graph.  Level~3
  Inductor flags control non-graph behaviors such as weight freezing and
  autotuning that cannot be expressed as graph rewrites.}
\label{fig:cand:fxpasslevel}
\end{figure}
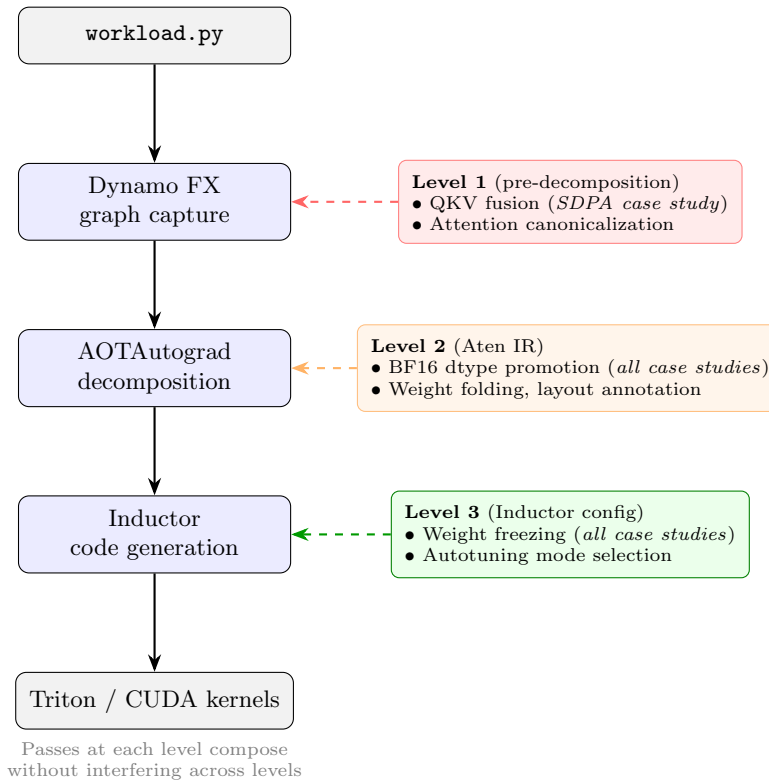

\end{document}